\documentclass[12pt]{article}

\usepackage{amsfonts,amssymb,amsthm,amsmath}
\usepackage[mathscr]{eucal}
\usepackage{hyperref}

\advance\textheight\topskip
\renewcommand{\tilde}{\widetilde}

\newcommand{\dd}{\partial}
\renewcommand{\d}{\partial}

\renewcommand{\geq}{\,{\geqslant}\,}
\renewcommand{\leq}{\,{\leqslant}\,}

\newcommand{\binner}[2]{%
  {\langle}\kern-4.15pt{\langle}#1{,}\,#2{\rangle}\kern-4.15pt{\rangle}}

\newcommand{\half}{\mathchoice{%
    \ffrac{1}{2}}{\frac{1}{2}}{\frac{1}{2}}{\frac{1}{2}}}

\newcommand{\ffrac}[2]{\raisebox{.5pt}%
  {\footnotesize$\displaystyle\frac{#1}{#2}$}\kern1pt}

\newcommand{\brst}{\mathsf{\Omega}}

\newcommand{\dl}[1]{\mathchoice{\ffrac{\dd}{\dd #1}}{\frac{\dd}{\dd
      #1}}{\ffrac{\dd}{\dd #1}}{\ffrac{\dd}{\dd #1}}}

\newcommand{\dover}[2]{\ffrac{\dd #1}{\dd #2}}
\newcommand{\ddd}[3]{\ffrac{\dd^2 #1}{\dd #2 \dd #3}}

\newcommand{\manifold}[1]{\mathscr{#1}}

\newcommand{\manM}{\manifold{M}}

\def\cG{\mathcal{G}}

\def\cL{\mathcal{L}}

\def\cN{\mathcal{N}}

\def\cS{\mathcal{S}}
\def\cT{\mathcal{T}}

\begin{document}
\pagestyle{myheadings}
\markboth{\textsc{\small Barnich, Bouatta, Grigoriev}}{%
  \textsc{\small Surface charges and Killing tensors for higher spin
    fields}}
\addtolength{\headsep}{4pt}


\begin{flushright}
  ULB-TH/05-11,\
  FIAN-TD/05-12,\\
  \texttt{hep-th/0507138}
\end{flushright}

\begin{centering}
  
  \vspace{1cm}
  
  \textbf{\Large{Surface charges and dynamical Killing tensors for
      higher spin gauge fields in constant curvature spaces}}
  
  \vspace{1.5cm}
  
  {\large G.~Barnich$^{a,*}$, N.~Bouatta$^{a,\dag}$ and
    M.~Grigoriev$^{b}$ }
  
  \vspace{1cm}

  \begin{minipage}{.8\textwidth}\small
    \mbox{}\kern-4pt$^a$Physique Th\'eorique et Math\'ematique,
    Universit\'e Libre de Bruxelles and International Solvay
    Institutes, Campus Plaine C.P. 231, B-1050 Bruxelles, Belgium
    
    \vspace{.5cm}
    
    \mbox{}\kern-4pt$^b$Tamm Theory Department, Lebedev Physics
    Institute, Leninsky prospect 53, 119991 Moscow, Russia
  \end{minipage}

\end{centering}

\vspace{1cm}

\begin{center}
  \begin{minipage}{.9\textwidth}
    \textsc{Abstract}. In the context of massless higher spin gauge
    fields in constant curvature spaces, we compute the surface
    charges which generalize the electric charge for spin one, the
    color charges in Yang-Mills theories and the energy-momentum and
    angular momentum for asymptotically flat gravitational fields. We
    show that there is a one-to-one map from surface charges onto
    divergence free Killing tensors. These Killing tensors are computed by
    relating them to a cohomology group of the first quantized BRST
    model underlying the Fronsdal action.
  \end{minipage}
\end{center}

\vfill

\noindent
\mbox{}
\raisebox{-3\baselineskip}{%
  \parbox{\textwidth}{ \mbox{}\hrulefill\\[-4pt]}} {\scriptsize$^*$
  Senior Research Associate of the National
  Fund for Scientific Research (Belgium).\\[-2pt]
  $^\dag$ Chercheur FRIA (Belgium).}

\thispagestyle{empty} \newpage



\section*{Introduction}

\addcontentsline{toc}{section}{Introduction}

In irreducible gauge theories, there are two kinds of charges. The
standard Noether charges for global symmetries are given by the
integral of the time component of a conserved current $j^\mu$,
$\partial_\mu j^\mu\approx 0$. Here, $\partial_\mu$ denotes the total
derivative and $\approx 0$ means zero ``on-shell'', i.e., on all
solutions of the field equations. More invariantly, in $n$ spacetime
dimensions, these conserved charges correspond to the integral over an
$n-1$ dimensional hypersurface of the $n-1$ form
$j^{n-1}=\sqrt{|g|}\frac{1}{(n-1)!}\epsilon_{\mu\mu_2\dots\mu_n}j^\mu
dx^{\mu_2}\dots dx^{\mu_n}$. The properties of the charges then follow
from Stokes theorem and on-shell closure of $j^{n-1}$,
$dj^{n-1}\approx 0$.

The second kind of charges are given by the integral over an $n-2$
dimensional hypersurface of an on-shell closed $n-2$ form $k^{n-2}$,
$dk^{n-2}\approx 0$. These charges are associated with gauge
symmetries, or more precisely, with ``reducibility parameters'', i.e.,
field dependent parameters of gauge symmetries restricted by the
condition that the associated gauge symmetries leave the fields
invariant on-shell. For electromagnetism, a constant gauge parameter
has this property, while for linearized gravity, examples of such
parameters which are field independent are provided by the Killing
vectors of the background metric.

It is important to realize that the surface charges of the linearized
theories are also relevant in non-abelian Yang-Mills theories or in
full gravity. Indeed, they can be used at infinity for configurations
that asymptotically approach some background so that the linearized
theory applies. A clear early discussion for massless spin 1 and spin
2 gauge fields has been given by Abbott and Deser
\cite{Abbott:1981ff,Abbott:1982jh}. These charges are also useful in
order to study physical properties of exact solutions to the full
interacting theory \cite{Iyer:1994ys,Wald:1999wa,Barnich:2003xg}.

A more systematic investigation of surface charges involves the
question when on-shell closed forms $k^{n-2}$ should be considered
trivial.  The natural answer is if they are on-shell exact,
$k^{n-2}\approx dl^{n-3}$. Surface charges are thus identified with
elements of the so-called characteristic cohomology group in form
degree $n-2$, $H^{n-2}_{\rm weak}(d)$, where ``weak'' means that
equalities are required to hold merely on-shell. Characteristic
cohomology has been discussed extensively in the mathematical
literature \cite{Vinogradov:1978,Tsujishita:1982aa,Vinogradov:1984,%
  Anderson1991,Bryant:1995}.

In regular Lagrangian gauge field theories, it has been shown
\cite{Barnich:1995db,Anderson:1996sc,Barnich:2000zw} that there is a
one-to-one map from $H^{n-2}_{\rm weak}(d)$ onto equivalence classes
of reducibility parameters. Reducibility parameters should be
considered equivalent when they coincide on-shell. Under some
technical assumptions, it has then been shown that irreducible gauge
theory do not admit characteristic cohomology in form degree lower
than $n-2$. It has also been shown explicitly in flat
\cite{Boulanger:2000rq} and in constant curvature spaces
\cite{Barnich:2004ts} of dimensions greater or equal to $3$ that the
non trivial reducibility parameters of massless spin 2 fields can
without loss of generality be taken not to depend on the fields and
thus reduce to the standard Killing vectors of the background metric.
Since a similar result can easily be established for the spin 1 case,
it then follows that for spin 1 and 2, the surface charges are
exhausted by those found by Abbott and Deser.

The aim of this paper is to generalize these results to the case of
massless higher spin gauge theories in constant curvature spaces in
the Fronsdal formulation \cite{Fronsdal:1978rb}
\cite{Fronsdal:1979vb}. As explained above, the surface charges for
the free theory will also be relevant for the full interacting theory
and exact solutions thereof, once a complete Lagrangian formulation
will be found (see the recent work in anti-De Sitter spaces by
Vasiliev \cite{Vasiliev:1990en,Vasiliev:1992av,Vasiliev:2003ev} for
progress in this direction).

In the first section, we briefly review the string inspired BRST first
quantized description of massless higher spin gauge fields both in
flat \cite{Ouvry:1986dv,Bengtsson:1986ys,Henneaux:1987cpbis} and in
(anti-) De Sitter spacetimes
\cite{Bengtsson:1990un,Bonelli:2003zu,Buchbinder:2001bs,Sagnotti:2003qa},
with or without trace constraints.
 
We then show in appendix A that the Cauchy order of these theories is
$2$ for spin $\geq 1$. According to general theorems
\cite{Barnich:1995db,Barnich:2000zw}, this implies in particular that
there is no characteristic cohomology in degree strictly lower than
$n-2$.

In the second section, the reducibility parameters for higher spin
gauge fields are related to dynamical Killing tensors that are
divergence free on-shell. The result of the appendix is then used to
show that the non trivial reducibility parameters of higher spin gauge
theories can be chosen independent of the fields in space-time
dimensions greater or equal to $3$. Hence, the classification of
surface charges for higher spin fields does not depend on the dynamics
and is exhausted by standard divergence free Killing tensors of the
constant curvature background. 

The Killing tensors are explicitly constructed in section 3. In flat
space, we use an investigation \cite{Barnich:2004cr} of the first
quantized model underlying the Fronsdal higher spin theories, where
the Killing tensor equations have been explicitly solved during the
computation of the cohomology group $H^{-3/2}(\brst)$. For
completeness, we re-derive these results in a form suitable for the
present context in appendix B. We then apply the standard procedure
consisting of inducing the results for constant curvature spaces from
the results of flat space in one dimension more. 

Both in anti-De Sitter \cite{Lopatin:1988hz,Vasiliev:2001wa} (see also
\cite{Bekaert:2005vh}) and in flat backgrounds \cite{Vasiliev:2005zu},
the divergence free Killing tensors constructed in this way are
parametrized by the same Young tableaux as the ``global higher spin
symmetries'' discussed originally in the context of the formulation of
higher spin gauge theories in terms of frame-like gauge fields and
generalized connections. This is no coincidence because global higher
spin symmetries leave invariant these fields, and thus also Fronsdal's
fields, which emerge as particular combinations thereof\footnote{The
  authors are grateful to M.~Vasiliev for pointing this out.}.

Finally, in the last section, we compute the explicit expressions of
the surface charges using general formulas \cite{Barnich:2001jy} valid
for generic irreducible gauge theories.

\section{Higher spin gauge fields in constant curvature spaces}
\label{sec:higher-spin-gauge}

The action for massless, double traceless fields
$\varphi_{\mu_1\dots\mu_s}$ of ``spin'' $s$ in constant curvature
spaces has been derived by Fronsdal \cite{Fronsdal:1979vb} in 4
dimensions. Its generalization in arbitrary dimensions has been
originally constructed in \cite{Lopatin:1988hz}. It is
invariant under the gauge transformations
\begin{eqnarray}
  \label{eq:1}
  \delta_\Lambda
  \varphi_{\mu_1\dots\mu_s}=s\nabla_{(\mu_1}\Lambda_{\mu_2\dots\mu_s)},
\end{eqnarray}
with $\Lambda_{\mu_2\dots\mu_s}$ a traceless gauge parameter. In
analogy with string field theory, the Fronsdal gauge field theory can
be compactly reformulated as the field theory associated to a BRST
first quantized relativistic particle with internal degrees of
freedom. In the case of non-vanishing constant curvature, we mostly
follow here the notations and conventions of Sagnotti and Tsulaia
\cite{Sagnotti:2003qa}, section 3, to which we refer for further
details.

The first quantized system consists of the conjugate variables
\begin{eqnarray}
  \label{eq:13}
  [x^\mu,p_\nu]=\imath \delta^\mu_\nu\ ,\quad
  [\alpha_1^\mu,\alpha_{-1}^\mu]=g^{\mu\nu}\ {\rm (bosonic)}, \quad
  [c_k,b_l]=\delta_{kl}\ {\rm (fermionic)}, 
\end{eqnarray}
where $\mu=0,\cdots,n-1$, $k,l=-1,0,1$ and $g^{\mu\nu}$ is the
inverse of the constant curvature metric $g_{\mu\nu}$ used to raise
and lower indexes.

The representation space for these commutation relations is chosen to
be the space of $x$ dependent wave functions taking values in the Fock
space for the oscillators. This Fock space is defined by
$b_0|0\rangle=0=c_1|{0}\rangle=b_1|0\rangle=\alpha_1^\mu|0\rangle$.
The space-time inner product involves the generally covariant volume
element $\sqrt{|g|}d^nx$, while the Fock space inner product is
defined by $\langle{0},c_0 0\rangle=1$. The coordinates $x^\mu$ act on
the states by multiplication and the action of $p_\mu$ on states is
defined according to
\begin{eqnarray}
  \label{eq:2}
  p_\mu=-\imath(\frac{\partial}{\partial
    x^\mu}-{\Gamma^\rho}_{\mu\nu}\alpha^\nu_{-1}\alpha_{1\rho}). 
\end{eqnarray}
The hermitian nilpotent BRST charge of the system is given by
\begin{multline}
  \label{eq:3}
  \brst = c_0(\tilde
  l_0-\frac{4}{L^2}N+\frac{6}{L^2})+c_1l_{-1}+c_{-1}l_1
  -c_{-1}c_1b_0-\\
  -\frac{6}{L^2}c_0c_{-1}b_1-\frac{6}{L^2}c_0b_{-1}c_1+
  \frac{4}{L^2}c_0c_{-1}b_1N
  +\frac{4}{L^2}c_0b_{-1}c_1N-\\
  -\frac{8}{L^2}c_0c_{-1}b_{-1}M+\frac{8}{L^2}c_0c_{1}b_{1}M^\dagger+
  \frac{12}{L^2}c_0c_{-1}b_{-1}c_1b_1,
\end{multline}
where
\begin{eqnarray}
  \label{eq:4}
l_{\pm 1}=\alpha_{\pm 1}\cdot p,\quad  [l_1,l_{-1}]=\tilde l_0,\quad
N=\alpha_{-1}\alpha_1+\frac{n}{2},\quad
M=\half \alpha_1\alpha_1,
\end{eqnarray}
and $L$ is the radius of anti-De Sitter space,
$\eta_{\alpha\beta}z^\alpha z^\beta\equiv z^2 =-L^2$, with
$\eta_{\alpha\beta}={\rm diag}(-1,+1,\dots,+1,-1)$. The results for de
Sitter space are then obtained by replacing $L^2$ by $-L^2$ in the
expression for the BRST charge, while the flat space results are
recovered by putting $\frac{1}{L^2}=0$.

Besides the BRST charge, there are three other operators relevant for
our discussion: the anti-hermitian ghost number operator defined by
\begin{eqnarray}
\cG=c_0b_0+c_1b_{-1}+c_{-1}b_{1}-\frac{1}{2},\label{eq:5}
\end{eqnarray}
with $[\cG,\brst]=\brst$; the hermitian total occupation number
operator,
\begin{eqnarray}
\cN_s=\alpha_{-1}\cdot\alpha_1+ c_{-1} b_1+b_{-1}c_1 -s,\label{eq:6}
\end{eqnarray}
with $[\cN_s,\brst]=0=[\cN_s,\cG]$ and finally, the trace operator
\begin{eqnarray}
\cT=\half
\alpha_1\alpha_1 + b_1c_1,\label{eq:7}
\end{eqnarray}
with $[\brst,\cT]=0=[\cG,\cT]$, $[\cN_s,\cT]=-2\cT$.

Let us associate to the most general ghost number $-1/2$ state
  \begin{multline}
  \label{eq:8}
  |\psi_{-1/2}\rangle=\int d^nx\ \Big(
  \frac{1}{s!}\alpha^{\mu_1}_{-1}\dots
  \alpha^{\mu_s}_{-1}\psi_{\mu_1\dots\mu_s}(x)~=
  \\
  +~\frac{-\imath}{(s-1)!}\alpha^{\mu_1}_{-1}\dots
  \alpha^{\mu_{s-1}}_{-1}c_0b_{-1} c_{\mu_1\dots\mu_{s-1}}(x)~+
  \\
  +~\frac{1}{(s-2)!}\alpha^{\mu_1}_{-1}\dots
  \alpha^{\mu_{s-2}}_{-1}c_{-1}b_{-1} d_{\mu_1\dots\mu_{s-2}}(x)\Big)
  |x\rangle|0\rangle
\end{multline}
of the Hilbert space the string field
\begin{multline}
  \label{eq:9}
  |\Psi_{-1/2}\rangle=\int d^nx\ \sum_{s} \Big(
  \frac{1}{s!}\alpha^{\mu_1}_{-1}\dots
  \alpha^{\mu_s}_{-1}\varphi_{\mu_1\dots\mu_s}(x)~+
\\
  + \frac{-\imath}{(s-1)!}\alpha^{\mu_1}_{-1}\dots
  \alpha^{\mu_{s-1}}_{-1}c_0b_{-1}
  C_{\mu_1\dots\mu_{s-1}}(x)~+
\\
  +~\frac{1}{(s-2)!}\alpha^{\mu_1}_{-1}\dots
  \alpha^{\mu_{s-2}}_{-1}c_{-1}b_{-1} D_{\mu_1\dots\mu_{s-2}}(x)\Big)
  |x\rangle|0\rangle\,,
\end{multline}
with $\varphi,C,D$ even bosonic and real tensor fields. If one now
imposes the constraints
$\cT|\Psi^T_{-1/2,s}\rangle=0=\cN_s|\Psi^T_{-1/2,s}\rangle $, $\cG
|\Psi^T_{-1/2,s}\rangle=-\half |\Psi^T_{-1/2,s}\rangle$, the Fronsdal
action for massless fields of spin $s$ in a constant curvature space
can be compactly written with the help of auxiliary fields as
\begin{eqnarray}
  \label{eq:9bis}
  S[\varphi,C,D]=-\half\langle \Psi^T_{-1/2,s},\brst
  \Psi^T_{-1/2,s}\rangle\ . 
\end{eqnarray}
If 
\begin{eqnarray}
  \label{eq:10}
  |\lambda_{-3/2}\rangle=\int d^nx
  \sum_s \Big(
\frac{\imath}{(s-1)!}\alpha^{\mu_1}_{-1}\dots\alpha^{\mu_{s-1}}_{-1}b_{-1}
\lambda_{\mu_1\dots\mu_{s-1}}(x)\Big)|x\rangle|0\rangle\ ,
\end{eqnarray}
is the general ghost number $-\frac{3}{2}$ state and 
\begin{eqnarray}
  \label{eq:11}
  |\Lambda_{-3/2}\rangle=\int d^nx
  \sum_s \Big(
\frac{\imath}{(s-1)!}\alpha^{\mu_1}_{-1}\dots\alpha^{\mu_{s-1}}_{-1}b_{-1}
\Lambda_{\mu_1\dots\mu_{s-1}}(x)\Big)|x\rangle|0\rangle,
\end{eqnarray}
an associated string field, with $\Lambda$ even bosonic and real
gauge parameters, the gauge transformations for this action can be
written as  
\begin{eqnarray}
  \label{eq:12}
  \delta_\Lambda |\Psi^T_{-1/2,s}\rangle=\brst |\Lambda^T_{-3/2,s}\rangle,
\end{eqnarray}
where
$\cT|\Lambda^T_{-3/2,s}\rangle=0=\cN_s|\Lambda^T_{-3/2,s}\rangle$, $\cG
|\Lambda^T_{-3/2,s}\rangle=-\frac{3}{2} |\Psi^T_{-1/2,s}\rangle$.

Removing the trace constraints on the string fields and the gauge
parameters, the action $S=-\half\langle \Psi_{-1/2,s},\brst
\Psi_{-1/2,s}\rangle$ continues to describe a perfectly consistent
gauge model in a constant curvature space, with gauge transformations
given by $\delta_\Lambda |\Psi_{-1/2,s}\rangle=\brst
|\Lambda_{-3/2,s}\rangle$, even though the model contains some
reducibility from the point of view of representations for different
values of $s$. More precisely, in $4$ dimensional flat space for
instance, the model without trace constraints describes massless
particles of helicities $-s, -s+2,\dots,s-2,s$. The trace constraints
project out the intermediate helicity states (see
e.g.~\cite{Henneaux:1987cpbis,Francia:2002pt,Barnich:2005ga} and
references therein).\footnote{How to describe massless particles of
  helicities $\pm s$ without imposing trace constraints is treated in
  \cite{Pashnev:1997rm,Pashnev:1998ti,Francia:2002aa,%
Francia:2002pt,Francia:2005bu}.
  Other aspects of higher spin fields in (A)dS spaces are discussed for
  instance in
  \cite{Deser:2001pe,Deser:2001us,Deser:2001xr,Deser:2004rr,%
Deser:2003gw,Deser:2004ji}. }

Removing the occupation number constraint, the action $S=-\half\langle
\Psi^{(T)}_{-1/2},\brst \Psi^{(T)}_{-1/2}\rangle$ describes the sum of
all these gauge models. In the $4$ dimensional flat case with trace
constraints, it describes massless particles of all helicities $\pm
s$ precisely once, while in the 4 dimensional anti-de Sitter case, it
describes singletons \cite{Fronsdal:1979vb}. 

Finally, removing the ghost number constraint and associating a string
field with all the states of the Hilbert space, the individual fields
being chosen real, of ghost number $-1/2$ {\em minus the ghost number
  of the corresponding state} and of parity this ghost number modulo 2,
$S=-\half\langle \Psi^{(T)}_{(s)},\brst \Psi^{(T)}_{(s)}\rangle$ is the
Batalin-Vilkovisky master action
\cite{Thorn:1987qj,Bochicchio:1987zj,Bochicchio:1987bd, Thorn:1989hm}
(see also \cite{Gaberdiel:1997ia,Barnich:2003wj}) for the gauge models
with or without trace constraints.

\section{From dynamical to standard Killing tensors}
\label{sec:from-dynam-stand}

In terms of the fields, the gauge transformations read
explicitly 
\begin{eqnarray}
\delta_\Lambda\varphi_{\mu_1\dots\mu_{s}} &=& s \nabla_{(\mu_1}
\Lambda_{\mu_2\dots\mu_{s})} \ , \nonumber \\
\delta_\Lambda D_{\mu_1\dots\mu_{s-2}} &=& \nabla^\mu
\Lambda_{\mu\mu_1\dots\mu_{s-2}} \ , \\
\delta_\Lambda C_{\mu_1\dots\mu_{s-1}} &=& \Big[\Box
+\frac{(s-1)(3-s-n)}{L^2}\Big]\Lambda_{\mu_1\dots\mu_{s-1}}
+\frac{(s-1)(s-2)}{L^2}
g_{(\mu_1\mu_2}}\Lambda^\prime_{\mu_3\dots\mu_{s-1)}\ .  
\nonumber\label{14}
\end{eqnarray}
where $\Lambda^\prime_{\mu_3\dots\mu_{s-1}}=g^{\mu_1\mu_2}
\Lambda_{\mu_1\mu_2\dots\mu_{s-1}}$.  Reducibility parameters are
defined by gauge parameters
$\Lambda_{\mu_1\dots\mu_{s-1}}=\Lambda_{\mu_1\dots\mu_{s-1}}[x,\varphi,C,D]$
that may depend on the fields and their derivatives and that satisfy 
\begin{equation}
 \label{15}
\begin{aligned}
  &\nabla_{(\mu_1} \Lambda_{\mu_2\dots\mu_{s})} \approx 0\,,
  \\
  &\nabla^\mu \Lambda_{\mu\mu_1\dots\mu_{s-2}} \approx 0 \,,
  \\
  & \Big[\Box
  +\frac{(s-1)(3-s-n)}{L^2}\Big]\Lambda_{\mu_1\dots\mu_{s-1}}
  +\frac{(s-1)(s-2)}{L^2}
  g_{(\mu_1\mu_2}\Lambda^\prime_{\mu_3\dots\mu_{s-1})} \approx 0\,.
 \end{aligned}
\end{equation}

In appendix A, we show that the fields $\varphi,C,D$ and their
derivatives can be split into two groups $(y_A,z_a)$ so that the
$y_A$ can be taken as coordinates on the surface defined by the field
equations and the $z_a$ as coordinates off this surface. Because
$H^{n-2}_{\rm weak}(d)$ is isomorphic to reducibility parameters, up
to weakly vanishing ones, non trivial reducibility parameters can be
chosen independent of the $z_a$,
$\Lambda_{\mu_1\dots\mu_{s-1}}=\Lambda_{\mu_1\dots\mu_{s-1}}(x,y)$.

We also show in appendix A that higher spin gauge theories are of
Cauchy order $2$, $\d_\mu y_A\subset y_A$, for $\mu\geq 2$. In
spacetime dimensions greater or equal to $3$, the first equation of
(\ref{15}) then implies that the reducibility parameters cannot depend $y_A$
either and are thus functions of $x^\mu$ alone,
$\Lambda_{\mu_1\dots\mu_{s-1}}= \Lambda_{\mu_1\dots\mu_{s-1}}(x)$. As
a consequence, the weak equalities can be replaced by strong ones in
equation (\ref{15}).  Furthermore, according to general results
\cite{Barnich:1995db,Barnich:2000zw}, theories of Cauchy order $2$
have trivial characteristic cohomology in form degrees strictly less
than $n-2$.

\section{Killing tensors in constant curvature spaces}
\label{sec:kill-tens-const}

\subsection{Flat space}
\label{sec:flat-space}

In flat space with $n\geq 3$, we have to solve the equations
\begin{eqnarray}
\d_{(\mu_1}\Lambda_{\mu_2\dots\mu_s)}= 0\ ,\quad
\d^\mu\Lambda_{\mu\mu_2\dots\mu_{s-1}} =  0\ 
, \quad 
\Box \Lambda_{\mu_1\mu_2\dots\mu_{s-1}} = 0\ , 
\label{16}
\end{eqnarray}
where the symmetric gauge parameters
$\Lambda_{\mu_1\mu_2\dots\mu_{s-1}}(x)$ are assumed to be traceless or
not, depending on whether the model with or without trace constraints
is considered.

These equations have been explicitly solved as a second step in the
computation of $H^{-3/2}(\brst)$ in section 4.4 of reference
\cite{Barnich:2004cr}\footnote{The ghost numbers of the first quantized
  model differ because, in reference \cite{Barnich:2004cr}, the
  non-Lagrangian context was emphasized and there was no need to have
  a fractional ghost number. Furthermore, what is called $y$ there
  corresponds to $x$ here.}.  This is no coincidence, but it is
related to the way the Batalin-Vilkovisky master action for higher
spin gauge fields is constructed out of the first quantized BRST
charge on the one hand and the relation between the characteristic
cohomology and the local BRST cohomology of the anti-field formalism
\cite{Barnich:1995db,Barnich:2000zw} on the other hand. General
results on relating field theory BRST cohomology groups to those of
the first quantized model will be given elsewhere. In order to be
self-contained, we will explicitly solve these equations again in
appendix B. In fact, we will solve them under slightly more general
assumptions since we allow for $\Lambda$ to depend smoothly on
$x^\mu$, contrary to reference \cite{Barnich:2004cr} where formal
power series were considered.

According to appendix B, the general solution to the first equation of
(\ref{16}) is given by 
\begin{eqnarray}
  \label{eq:14}
  \Lambda_{\mu_1\dots\mu_{s-1}}(x)= \sum^{s-1}_{m=0}
A_{\mu_1\cdots\mu_{s-1}|\nu_1\cdots\nu_m }x^{\nu_1}\cdots
x^{\nu_m} \ , \label{solf1}
\end{eqnarray}
where the constant coefficients $A_{\mu_1\cdots\mu_{s-1}|\nu_1\cdots\nu_m }$
are required to have the symmetries of the Young tableaux
\begin{equation} \label{eq:YT2}
  \begin{picture}(100,25)
    \put(0,-10){
      \put(-12,23){\tiny $\mu$}
      \put(-12,13){\tiny $\nu$}
      \put(0,30){\line(1,0){90}}
      \put(0,20){\line(1,0){90}}
      \put(90,20){\line(0,1){10}}
      \put(80,20){\line(0,1){10}}
      \put(63,21.5){$\cdots$}
      \put(60,20){\line(0,1){10}}
      \put(50,10){\line(0,1){20}}
      \put(0,10){\line(1,0){50}}
      \put(40,10){\line(0,1){20}}
      \put(16.5,21.5){$\cdots$}
      \put(16.5,11.5){$\cdots$}
      \put(10,10){\line(0,1){20}}
      \put(0,10){\line(0,1){20}}
    }
  \end{picture}\ .
\end{equation}
The second equation of (\ref{16}) then requires these tableaux to be
traceless in the columns, while the last equation arises as a
compatibility condition for the first two and requires the tableaux to
be traceless in the second row. Finally, if the trace constraint on the
gauge parameter is imposed, the tableaux are required to be traceless
in the first row as well, and thus to be completely traceless.

\subsection{Curved space}
\label{sec:curved-space}

Let $\manM$ be De Sitter or anti-De Sitter space, defined through the
embedding $z^2\equiv\eta_{\alpha\beta}z^\alpha z^\beta=\varepsilon
L^2$ in flat $n+1$ dimensional space $M^{n+1}$, with
$\varepsilon=1$ for De Sitter and $\varepsilon=-1$ for anti-De Sitter
space respectively. In describing tensor fields on $(A)dS$ we closely
follow the approach of~\cite{Fronsdal:1979vb}.  Take functions $x^\mu$
on $M^{n+1}$ such that $z^\alpha\dl{z^\alpha}x^\mu=0$ and
$x^\mu,r=\sqrt{\varepsilon z^2}$ can be taken as local coordinates on
$M^{n+1}$ (minus the origin). It follows that, restricted to
$\manM$, the functions $x^\mu$ are local coordinates on $\manM$.
Conversely, any admissible local coordinates ${x^\prime}^\mu$ on
$\manM$ can be lifted to such coordinate functions on $M^{n+1}$.
These definitions imply that the invertible change of coordinates
$z^\alpha=z^\alpha(r,x^\mu)$ satisfies
\begin{align}
z_A\dover{z^A}{x^\mu}&=0\,,
\qquad&
 \dover{x^\mu}{z^A}\dover{z^A}{x^\nu}&=\delta^\mu_\nu\,,
\\
 \dover{z^A}{r}&=\frac{z^A}{r}\,,
\qquad&
 \dover{z^A}{x^\mu}\dover{x^\mu}{z^{B}}+\varepsilon
\frac{z^A z_{B}}{r^2}&=\delta^A_{B}\,. 
\end{align}
In addition to the embedding map $\manM\to M^{n+1}$, one defines the
projection $M^{n+1}\to \manM$ which sends a point with coordinates
$r,x^\mu$ to a point with coordinates $x^\mu,L$. Using the projection
one can establish a one-to-one correspondence between covariant tensor
fields on $\manM$ and their images on $M^{n+1}$ under the pullback
of the projection map. It can be useful to consider also the
modified map given by 
\begin{eqnarray}
  \label{eq:21}
  \Lambda_{A_1\dots A_{s-1}}(z)=
\left(\frac{\sqrt{\varepsilon z^2}}{L}\right)^N\dover{x^{\mu_1}_L}{z^{A_1}}\dots
\dover{x^{\mu_{s-1}}_L}{z^{A_{s-1}}}
  \Lambda_{\mu_1\dots\mu_{s-1}}(x(z))\,.
\end{eqnarray}
where $\dover{x^\mu_L}{z^A}$ is $\dover{x^\mu}{z^A}$ evaluated at $r=L$.
For $N=-s+1$ this coincides with the pullback
of the projection map, which can be seen using
\begin{equation}
z^A\dl{z^A}(\dover{x^\mu}{z^{B}})=r\dl{r}(\dover{x^\mu}{z^{B}})=
-\dover{x^\mu}{z^{B}}
\end{equation}
which in turn implies
$\dover{x^\mu}{z^A}=(\frac{r}{L})^{-1}\dover{x^\mu_L}{z^A}$.  The
tensor $\Lambda_{A_1\dots A_{s-1}}(z)$ defined by~\eqref{eq:21}
satisfies
\begin{eqnarray}
  \label{eq:22}
  z^{A_i}\Lambda_{A_1\dots A_i\dots {A}_{s-1}}(z)=0\ ,
 \   z^{B} \frac{\dd}{\dd
    z^{B}}\Lambda_{A_1\dots {A}_i\dots {A}_{s-1}}(z) =N
\Lambda_{A_1\dots {A}_i\dots {A}_{s-1}} (z) \,.
\end{eqnarray}
There is a one-to-one map from covariant tensor fields on $\manM$
onto those on $M^{n+1}$ satisfying~\eqref{eq:22}. In terms of
components the inverse map is given by
\begin{eqnarray}
  \label{eq:27}
  \Lambda_{\mu_1\dots\mu_{s-1}}(x)=\dover{z^{A_1}_L}{x^\mu_1}\dots
\dover{z^{A_{s-1}}_L}{x^{\mu_{s-1}}}
\Lambda_{A_1\dots  {A}_{s-1}}(z_L)\ .
\end{eqnarray}
where $z_L^A=z^A|_{r=L}$. Note also that $\frac{z^A}{r}=\frac{z^A_L}{L}$.

For $N=s-1$, the definition (\ref{eq:21}) implies 
\begin{eqnarray}
  \label{eq:17}
  \dd_{(A_1}\Lambda_{A_2\dots A_s)}(z)=
(\frac{{\varepsilon z^2}}{L^2})^{s-1}\dover{x^{\mu_1}}
{z^{A_1}}\dots\dover{x^{\mu_s}}{z^{A_s}}
\nabla_{(\mu_1}\Lambda_{\mu_2\dots\mu_s)}(x) \,.
\end{eqnarray}
as can be seen by direct computation using the fact that the Levi-Civita
connection on $\manM$ is given in terms of the embedding by
\begin{eqnarray}
  \label{eq:24}
  \Gamma^\rho_{\mu\nu}(x)=\dover{x^\rho_L}{z^B} 
\dover{{}^2z^B_L}{x^\mu\dd x^\nu}=
\dover{x^\rho}{z^B} \dover{{}^2z^B}{x^\mu\dd
  x^\nu}\,.
\end{eqnarray}
Hence, 
\begin{eqnarray}
  \label{eq:30}
  \nabla_{(\mu_1}\Lambda_{\mu_2\dots\mu_s)}(x)=0\label{eq:curv}
\end{eqnarray}
implies 
\begin{eqnarray}
  \label{eq:25}
  \dd_{(A_1}\Lambda_{A_2\dots {A}_s)}(z)=0\,.
\end{eqnarray}
According to the previous subsection, the
general solution to (\ref{eq:25}) is given by
\begin{eqnarray}
  \label{eq:26}
  \Lambda_{A_1\dots {A}_{s-1}}(z)=\sum_{m=0}^{s-1}A_{A_1\dots {A}_{s-1}| 
{B}_1\dots{B}_m}z^{{B}_1}\dots z^{{B}_m}\ ,
\end{eqnarray}
where the constant coefficients
$A_{A_1\dots {A}_{s-1}|{B}_1\dots{B}_m}$ have again the
symmetries of two row Young tableaux, the indexes running now over
$n+1$ values. Only the last term with $m=s-1$ and involving, for
$s>1$, a Young tableau with two rows of equal length $s-1$,
\begin{equation} \label{eq:YT4}
  \begin{picture}(100,25)
    \put(0,-10){
      \put(-12,23){\tiny $A$}
      \put(-12,13){\tiny ${B}$}
      \put(0,30){\line(1,0){90}}
      \put(0,20){\line(1,0){90}}
      \put(80,10){\line(0,1){20}}
      \put(0,10){\line(1,0){90}}
      \put(90,10){\line(0,1){20}}
      \put(36.5,21.5){$\cdots$}
      \put(36.5,11.5){$\cdots$}
      \put(10,10){\line(0,1){20}}
      \put(0,10){\line(0,1){20}}
    }
  \end{picture}\ ,
\end{equation}
satisfies the conditions (\ref{eq:22}) with $N=s-1$. The solutions to
$\nabla_{(\mu_1}\Lambda_{\mu_2\dots\mu_s)}(x)=0$ are thus explicitly
given by
\begin{eqnarray}
  \label{eq:28}
  \Lambda_{\mu_1\dots\mu_{s-1}}(x)
=\dover{z^{A_1}_L}{x^\mu_1}\dots
\dover{z^{A_{s-1}}_L}{x^{\mu_{s-1}}} A_{A_1\dots {A}_{s-1}| 
{B}_1\dots{B}_{s-1}}z^{{B}_1}_L\dots z^{{B}_{s-1}}_L.
\end{eqnarray}
and are characterized by a {\em single} 2 row rectangular Young
tableau of length $s-1$ in $n+1$ dimensions.

Using 
\begin{eqnarray}
  \label{eq:18}
  \ddd{x^\mu}{z^A}{z^B}=-\dover{x^\mu}{z^A}\frac{\varepsilon
    z_B}{r^2}-\dover{x^\mu}{z^B}\frac{\varepsilon
    z_A}{r^2}
  -\dover{x^\mu}{z^C}\dover{x^\nu}{z^A}\dover{x^\lambda}{z^B} 
\ddd{z^C}{x^\nu}{x^\lambda}, 
\end{eqnarray}
definition (\ref{eq:21}) for $N=s-1$ implies
\begin{multline}
  \label{eq:19}
  \d^A\Lambda_{AA_2\dots A_{s-1}}=(\frac{\varepsilon
    z^2}{L^2})^{s-2}\Big[\dover{x^{\mu_2}}{z^{A_2}}\dots
\dover{x^{\mu_{s-1}}}{z^{A_{s-1}}}\nabla^\mu
\Lambda_{\mu\mu_2\dots\mu_{s-1}}~-
\\
-~(\frac{\varepsilon z_{A_2}}{r^2}\dover{x^{\mu_3}}{z^{A_3}}\dots
\dover{x^{\mu_{s-1}}}{z^{A_{s-1}}}+\dover{x^{\mu_3}}{z^{A_2}}
\frac{\varepsilon z_{A_3}}{r^2}\dover{x^{\mu_4}}{z^{A_4}}\dots
\dover{x^{\mu_{s-1}}}{z^{A_{s-1}}}+\dots~+
\\
+~\dover{x^{\mu_3}}{z^{A_2}}\dots
\dover{x^{\mu_{s-1}}}{z^{A_{s-2}}}\frac{\varepsilon
  z_{A_{s-1}}}{r^2})
\Lambda^\prime_{\mu_3\dots\mu_{s-1}}\Big].
\end{multline}
By contracting with $\eta^{A_1A_p}$ and using
$g^{\mu_1\mu_2}=\dover{x^{\mu_1}_L}{z^{A_1}}\eta^{A_1A_2}
\dover{x^{\mu_{2}}_L}{z^{A_2}}$, definition (\ref{eq:21}) for $N=s-1$
also implies
\begin{eqnarray}
  \label{eq:23}
  \Lambda^A_{A_2\dots A_{p-1}AA_{p+1}\dots
  A_{s-1}}=(\frac{\varepsilon
    z^2}{L^2})^{s-2}\dover{x^{\mu_3}}{z^{A_2}}
\dots \dover{x^{\mu_{p}}}{z^{A_{p-1}}}
\dover{x^{\mu_{p+1}}}{z^{A_{p+1}}}\dots
\dover{x^{\mu_{s-1}}}{z^{A_{s-1}}}\Lambda^\prime_{\mu_3\dots\mu_{s-1}}\,.
\end{eqnarray}
It follows that 
the equations 
\begin{eqnarray}
  \label{eq:31}
  \nabla^\mu
\Lambda_{\mu\mu_2\dots\mu_{s-1}}=0
\end{eqnarray}
are equivalent to
\begin{equation}
z^2\d^A\Lambda_{AA_2\dots
  {A}_{s-1}}+\sum_{p=2}^{s-1} z_{A_p}
\Lambda^A_{A_2\dots A_{p-1}AA_{p+1}\dots
  A_{s-1}}=0\,. \label{eq:AA}
\end{equation}
This equation imposes the following constraint on the rectangular 2
row Young tableau:
\begin{multline}
  \label{eq:29}
  \sum_{\sigma\in \cS(s)} \Big[(s-1)
  \eta_{B_{\sigma(1)}B_{\sigma(2)}}
A^A_{A_2\dots A_{s-1}|A B_{\sigma(3)}\dots B_{\sigma(s)}}
~+
\\
~+
\sum_{p=2}^{s-1}\eta_{A_p B_{\sigma(1)}} A^\prime_{A_2\dots A_{p-1}A_{p+1}\dots
  A_{s-1}|B_{\sigma(2)}B_{\sigma(3)}\dots B_{\sigma(s)}}\Big]=0\,.
\end{multline}

The equation 
\begin{eqnarray}
  \label{eq:16}
  \Big[\Box
  +\frac{(s-1)(3-s-n)}{L^2}\Big]\Lambda_{\mu_1\mu_2\dots\mu_{s-1}}
+\frac{(s-1)(s-2)}{L^2}
g_{(\mu_1\mu_2}\Lambda^\prime_{\mu_3\dots\mu_{s-1})} =0\ ,
\end{eqnarray}
is then automatically satisfied since it is the compatibility
condition for \eqref{eq:curv} and \eqref{eq:31}. 

If the trace condition is imposed to start with, the problem
simplifies since the rectangular 2 row
Young tableau is required to be traceless in the first row from the
start. Equations \eqref{eq:AA} then simply require this Young tableau to be
traceless in the columns as well, while the compatibility
condition reduces, as in the flat case, to requiring the tableau to be
traceless in the second row, and thus to be completely traceless. 

\section{Surface charges for higher spin gauge fields}
\label{sec:surf-charg-high}

The Euler-Lagrange derivatives associated to action (\ref{eq:9bis})
are explicitly given by 
\begin{multline}
\cL_{\varphi}^{\mu_1\dots\mu_s}=\frac{\sqrt{|g|}}{s!}\Big[
\Box\varphi^{\mu_1\dots\mu_s}-s\nabla^{(\mu_1}C^{\mu_2\dots\mu_{s})}
-\frac{1}{L^2}\big[4s(s-1)g^{(\mu_1\mu_2}D^{\mu_3\dots\mu_s)}
-
\\
-s(s-1)
g^{(\mu_1\mu_2}\varphi^{\prime\mu_3\dots\mu_s)}+
[(2-s)(3-n-s)-s]\varphi^{\mu_1\dots\mu_s}\big]\Big], 
\end{multline}
\begin{multline}
\cL_{C}^{\mu_1\dots\mu_{s-1}}=\frac{\sqrt{|g|}}{(s-1)!}
\Big[-C^{\mu_1\dots\mu_{s-1}}+\nabla_\mu
\varphi^{\mu\mu_1\dots\mu_{s-1}}-(s-1)
\nabla^{(\mu_1}D^{\mu_2\dots\mu_{s-1})}\Big],
\end{multline}
\begin{multline}
  \cL_{D}^{\mu_1\dots\mu_{s-2}}=\frac{\sqrt{|g|}}{(s-2)!}\Big[-\Box
  D^{\mu_1\dots\mu_{s-2}}
  +\nabla_{\mu}C^{\mu\mu_1\dots\mu_{s-1}}-\frac{1}{L^2}\big[4
\varphi^{\prime\mu_1\dots\mu_{s-2}} +
  \\
  +(s-2)(s-3) g^{(\mu_1\mu_2}D^{\prime\mu_3\dots\mu_{s-2})}-
  [s(n+s-2)+6]D^{\mu_1\dots\mu_{s-2}}\big]\Big],
\end{multline}
Using the Noether identities
\begin{multline}
-s\nabla_\mu\cL_{\varphi}^{\mu\mu_1\dots\mu_{s-1}}-\nabla^{(\mu_1}
  \cL_{D}^{\mu_2\dots\mu_{s-1})}+\big[\Box
+\frac{(s-1)(3-s-n)}{L^2}\big]\cL_{C}^{\mu_1\dots\mu_{s-1}}+\\
+\frac{(s-1)(s-2)}{L^2}g^{(\mu_1\mu_2}
\cL_{C}^{\prime \mu_3\dots\mu_{s-1})}=0\,,
\end{multline}
the weakly vanishing Noether current $j^\beta_\Lambda$ defined by 
\begin{multline}
s\cL_{\varphi}^{\mu_1\dots\mu_s} \nabla_{(\mu_1}
\Lambda_{\mu_2\dots\mu_{s})} +\cL_{D}^{\mu_1\dots\mu_{s-2}} \nabla^\mu
\Lambda_{\mu\mu_1\dots\mu_{s-2}} + \\+
\cL_{C}^{\mu_1\dots\mu_{s-1}}\Big[\big[\Box
+\frac{(s-1)(3-s-n)}{L^2}\big]\Lambda_{\mu_1\dots\mu_{s-1}}
+\\+\frac{(s-1)(s-2)}{L^2}
g_{(\mu_1\mu_2}\Lambda^\prime_{\mu_3\dots\mu_{s-1})}\Big]
=\partial_\beta j^\beta_\Lambda,
\end{multline}
is found to be 
\begin{multline}
j^\beta_\Lambda=s\cL_{\varphi}^{\beta\mu_1\dots\mu_{s-1}}
\Lambda_{\mu_1\dots\mu_{s-1}}+\cL_{D}^{\mu_2\dots\mu_{s-1}}
\Lambda^\beta_{\mu_2\dots\mu_{s-1}}+\\+\cL_{C}^{\mu_1\dots\mu_{s-1}}
\nabla^\beta\Lambda_{\mu_1\dots\mu_{s-1}}-
\nabla^\beta\cL_{C}^{\mu_1\dots\mu_{s-1}}
\Lambda_{\mu_1\dots\mu_{s-1}}\,.
\end{multline}
Applying previous results derived for generic gauge theories
\cite{Barnich:2001jy}, non trivial on-shell conserved $n-2$-forms are
associated with non trivial reducibility parameters and thus with
divergence free Killing tensors $\Lambda_{\mu_1\dots\mu_{s-1}}(x)$.
Explicitly, the $n-2$ forms $k_\Lambda=k^{[\alpha\beta]}_\Lambda
(d^{n-2}x)_{\alpha\beta}$, with
$(d^{n-2}x)_{\alpha\beta}=\frac{1}{2(n-2)!}\epsilon_{\alpha\beta
\mu_3\dots\mu_{n}}
dx^{\mu_3}\dots dx^{\mu_n}$, $\epsilon_{01\dots n-1}=1$, satisfy
\begin{equation}
\d_\alpha k^{[\alpha\beta]}_\Lambda=j^\beta_\Lambda
\end{equation}
and are constructed out of $j^\beta_\Lambda$ according to 
\begin{eqnarray}
k^{[\alpha\beta]}_\Lambda 
= \frac{1}{2}\phi^i\frac{\d j^\beta_\Lambda}{\d\phi^i_\alpha}
+ (\frac{2}{3}\phi^i_\lambda -
\frac{1}{3}\phi^i\d_\lambda)\frac{\d^S
j^\beta_\Lambda}{\d\phi^i_{\lambda\alpha}} -
(\alpha\longleftrightarrow\beta) \ , 
\end{eqnarray}
where
$\phi^i\equiv(\varphi_{\mu_1\dots\mu_s},C_{\mu_1\dots\mu_{s-1}},
D_{\mu_1\dots\mu_{s-2}})$, the subscripts on $\phi^i$ denoting
derivatives and $\partial^S {\phi^j_{\alpha \gamma}}/
\partial {\phi^i_{\delta \beta}}
=\delta^j_i\delta^\delta_{(\alpha}\delta^\beta_{\gamma)}$.
Direct computation gives
\begin{multline}\label{eq:n-2}
k^{[\alpha\beta]}_\Lambda = \frac{\sqrt{|g|}}{(s-1)!} [\nabla^\alpha
\varphi^{\beta\mu_1\dots\mu_{s-1}}\Lambda_{\mu_1\dots\mu_{s-1}} + (s-1)
\nabla^\beta D^{\mu_1\dots\mu_{s-2}}\Lambda^\alpha_{\mu_1\dots\mu_{s-2}} 
 +\\ 
+ \varphi^{\alpha\mu_1\dots\mu_{s-1}}\nabla^\beta 
\Lambda_{\mu_1\dots\mu_{s-1}}
+ (s-1) D^{\mu_1\dots\mu_{s-2}}\nabla^\alpha
\Lambda^\beta_{\mu_1\dots\mu_{s-2}}  +\\
+(s-1)C^{\alpha\mu_1\dots\mu_{s-2}}\Lambda^\beta_{\mu_1\dots\mu_{s-2}}-
(\alpha\longleftrightarrow\beta)] \ .
\end{multline}
{\bf Comments:} 

(i) The fields $C_{\mu_1\dots\mu_{s-1}}$ are auxiliary
since they can be eliminated by their own equations of motion,
$\cL_C^{\mu_1\dots\mu_{s-1}}=0$. If this is done on the level of
$j^\beta_\Lambda$ before the computation of
$k^{[\alpha\beta]}_\Lambda$, both expressions coincide only after
using that $\Lambda_{\mu_1\dots\mu_{s-1}}(x)$ are divergence free
Killing tensors.

(ii) The conserved $n-2$ forms of the Fronsdal theory for higher spin gauge
fields are obtained by substituting $D=\half\varphi^\prime$ in
$k_\Lambda$, with $\varphi$ taken to be double traceless and $\Lambda$
traceless.

(iii) After these steps, the $n-2$ forms $k_\Lambda$ for the spin $1$ and
$2$ cases can easily be seen to coincide with the Abbott and Deser
expressions in the form given in \cite{Barnich:2001jy} section 6.

\section{Conclusion and perspectives}

In this paper we have computed the non trivial surface charges for
higher spin gauge fields in constant curvature spacetimes by relating
them to divergence free dynamical Killing tensors. In the anti-De
Sitter case where definite progress towards constructing non trivial
interactions has been made (see e.g.~\cite{Bekaert:2005vh} for a
review), it remains to be seen if the surface charges for higher spin
fields can play as prominent a role as the Abbott-Deser charges do in
the case of asymptotically anti-De Sitter gravity.

It is quite intriguing to realize that the computation of a physically
relevant cohomology group in the field theory effectively reduces to
the computation of a cohomology group of the first quantized model: in
this article, we have shown in particular that the surface charges, or
more precisely, that the characteristic cohomology group $H^{n-2}_{\rm
  weak}(d)$ in the field theory is isomorphic to the BRST cohomology
group $H^{-3/2}(\brst)$ of the first quantized particle model
underlying the theory of massless higher spin gauge fields. In other
words, for the cohomology groups $H^{-3/2}(\brst)$ for which no clear
physical interpretation seems to have been known (see e.g.
chapter 11.1.2 of \cite{Henneaux:1992ig}), we have been able to find
such an interpretation in the associated field theory.

A reason why the cohomology group $H^{-3/2}(\brst)$ has not
attracted more interest so far is that this cohomology is normally 
argued to vanish \cite{Henneaux:1987cpbis}. As usual in such cases,
this is because the cohomology of $\brst$ is computed in different
spaces: in \cite{Henneaux:1987cpbis}, the cohomology is computed in
momentum space at non zero momentum, whereas in \cite{Barnich:2004cr},
it is computed in $x$-space and is described by polynomials in
$x$, which means that it is concentrated precisely at zero
momentum\footnote{G.B.~is grateful to G.~Bonelli for a useful
  discussion on this point.}.  

A similar analysis can of course be repeated in the BRST formulation
of string theory. In this way, one should be able to construct
meaningful surfaces charges in string field theory associated with
negative ghost number cohomology classes of the BRST charge in the
zero momentum sector. We plan to analyze this question in more details
in the near future.

\section*{Acknowledgments}

\addcontentsline{toc}{section}{Acknowledgments}

Useful discussions with X.~Bekaert, G.~Bonelli, G.~Comp\`ere,
A.~Semikhatov and I.~Tipunin are gratefully acknowledged.  The work of
G.B.~and N.B.~is supported in part by a ``P{\^o}le d'Attraction
Interuniversitaire'' (Belgium), by IISN-Belgium, convention 4.4505.86,
by Proyectos FONDECYT 1970151 and 7960001 (Chile) and by the European
Commission program MRTN-CT-2004-005104, in which these authors are
associated to V.U.~Brussels. The work of M.G. is supported by~RFBR
grant 04-01-00303 and by the grant LSS-1578.2003.2.

{\bf Note added:} While completing this work, reference
\cite{Bekaert:2005ka} appeared where, in particular, the dynamical
Killing tensor equations in flat spacetimes have been solved under
simplifying assumptions. We also understand that the standard Killing
tensor equations in constant curvature spaces have been solved
previously in \cite{Thompson:1986} and that the connection with
rectangular 2 row Young tableaux of length $s-1$ in $n+1$
dimensions, both in flat and in constant curvature spaces, has been
made previously in \cite{McLen:2004}.

\appendix 

\section{Cauchy order of higher spin gauge fields}

Because of gauge invariance, the (left hand sides of the) equations of
motion are not all independent functions on the space where the fields
and their derivatives are considered as independent coordinates: they
satisfy the so-called Noether identities. As a consequence, we can
split the equations of motion into two groups, the ``independent
equations'' $L_a$ and ``the dependent equations'' $L_\Delta$ (the
dependent equations hold as a consequences of the independent ones:
$L_a = 0$ implies $L_\Delta = 0$). We will show below that the fields
and their derivatives can be split into two groups, ``independent
coordinates'' $y_A$, which are not constrained by the equations of
motions, and ``dependent coordinates'' $z_a$, which are such that
$L_a=0$ can be solved for $z_a$ in terms of the $y_A$. This implies in
particular that the gauge transformations (\ref{14}) provide a
generating set of irreducible gauge transformations.

Without loss of generality we can focus in the following on flat
space. Indeed, the splits of equations and variables in flat space
will also be valid in spaces with non vanishing constant curvature
because the terms with the highest number of derivatives in the
Noether identities and in the equations of motions in constant
curvature spaces are the same in the two cases. We will start
by discussing the model without trace constraint.

The equations of motion in flat space are explicitly given by 
\begin{eqnarray}\label{eq:flati}
\Box \varphi_{\mu_1\dots\mu_s} = s\d_{(\mu_1} C_{\mu_2\dots\mu_s)} \ ,
\nonumber\\ 
\d^\mu\varphi_{\mu\mu_2\dots\mu_s}-(s-1)\d_{(\mu_2} D_{\mu_3\dots\mu_s)}
= C_{\mu_2\dots\mu_s}  \ , \\
\Box D_{\mu_1\dots\mu_{s-2}} = \d^\mu C_{\mu\mu_1\dots\mu_{s-2}} \ , \nonumber
\end{eqnarray}
and the corresponding gauge transformations by 
\begin{eqnarray}
\delta_\Lambda\varphi_{\mu_1\dots\mu_{s}} &=& s \d_{(\mu_1}
\Lambda_{\mu_2\dots\mu_{s})} \ , \nonumber \\
\delta_\Lambda D_{\mu_1\dots\mu_{s-2}} &=& \d^\mu
\Lambda_{\mu\mu_1\dots\mu_{s-2}} \ , \\
\delta_\Lambda C_{\mu_1\dots\mu_{s-1}} &=& \Box
\Lambda_{\mu_1\dots\mu_{s-1}} \ . 
\nonumber
\end{eqnarray}
In terms of the Euler-Lagrange derivatives
$\cL_{\varphi}^{\mu_1\dots\mu_s}$, $\cL_{C}^{\mu_1\dots\mu_{s-1}}$
$\cL_{D}^{\mu_1\dots\mu_{s-2}}$ associated to action \eqref{eq:9bis},
the Noether identities are given by
\begin{eqnarray}
s\,\d_\mu \cL_{\varphi}^{\mu\mu_2\dots\mu_{s}}-
\Box\cL^{\mu_1\dots\mu_{s-1}}_C + \d^{(\mu_2}\cL^{\mu_3\dots\mu_s)}_D = 0 \ .
\end{eqnarray}
These identities can be solved for $\d_0\cL^{0\mu_2\dots\mu_s}_\varphi$
and their derivatives, which constitute the $L_\Delta$. The remaining
equations and their derivatives are the $L_a$, they are all
independent because they can be solved for the $z_a$ listed in the
table below. The remaining fields and their derivatives are the $y_A$.
We have separated the tensor indexes on the fields from the indexes
indicating derivatives by a vertical bar. The Greek indexes go from
$0$ to $n-1$, the Latin indexes from $1$ to $n-1$, while the
barred Latin indexes go from $2$ to $n-1$. The multi-indexes
$(\mu),(k),(\nu),(\lambda)$ are of orders $|\mu|=s=(k),
|\nu|=s-1,|\lambda|=s-2$, while the indexes
$(\alpha),(\beta),(m),(\bar l),(\gamma),(n),(\delta)$ 
are of orders 
$|\alpha|=k, |\beta|=k-1=|m|=|\bar l|, |\gamma|=k-2=|n|,|\delta|=k-3$. 
The listed $y_A$ have the property we
wanted to show, namely $\partial_{\bar l}y_A\subset y_A$.

{\footnotesize
\begin{eqnarray} \begin{array}{l|l|l|l|l}  & \cL_\Delta & 
\cL_a & z_a & y_A
\\
\hline\rule{0em}{3ex} V^0 & \o & \o & \o & \varphi_{(\mu)},
C_{(\nu)}, D_{(\lambda)}
\\
\hline\rule{0em}{3ex} V^1 & \o & \cL^{(\nu)}_C & \varphi_{0(\nu)|0} &
\varphi_{(k)|0}, \varphi_{(\mu)|i}, C_{(\nu)|\alpha}, 
\\ &&&&D_{(\lambda)|\alpha}\\
\hline\rule{0em}{3ex} V^2 & \o & \cL^{(\mu)}_\varphi, \cL^{(\lambda)}_D, 
\d_\beta\cL^{(\nu)}_C &
\varphi_{0(\nu)|0\beta}, \varphi_{(k)|00}, \varphi_{0(\nu)|11},
 & \varphi_{(k)|m\rho}, \varphi_{0(\nu)|\bar{l}p}, C_{(\nu)|\alpha_1\alpha_2},
\\
&&& D_{(\lambda)|00}&D_{(\lambda)|m\rho}
\\
\hline\rule{0em}{3ex} V^3 & \d_0\cL_\varphi^{0(\nu)} &
\d_{\beta_1\beta_2}\cL^{(\nu)}_C, \d_\gamma\cL^{(k)}_\varphi, &
\varphi_{0(\nu)|0\beta_1\beta_2}, \varphi_{(k)|00\gamma},
\varphi_{0(\nu)|11n}, &
\varphi_{(k)|m_1m_2\rho}, \varphi_{0(\nu)|\bar{l_1}\bar{l_2}p},
C_{(\nu)|\alpha_1\alpha_2\alpha_3},
\\&&\d_m\cL_\varphi^{0(\nu)},\d_\gamma\cL_D^{(\lambda)} &
D_{(\lambda)|00\gamma} &  
D_{(\lambda)|m_1m_2\rho}
\\
\hline\rule{0em}{3ex}.
\\
.
\\
.
\\
\hline\rule{0em}{3ex}
 V^k & \d_{(\delta)0}\cL^{0(\nu)} & \d_{(\beta)}\cL^{(\nu)}_C,
\d_{(\gamma)}\cL_\varphi^{(k)},  & \varphi_{0(\nu)|0(\beta)},
\varphi_{(k)|00(\gamma)}  & \varphi_{(k)|(m)\rho},
\varphi_{0(\nu)|(\bar{l})p},
\\&&\d_{(n)}\cL_\varphi^{0(\nu)},
\d_{(\gamma)}\cL^{(\lambda)}_D & \varphi_{0(\nu)|11(n)},
D_{(\lambda)|00(\gamma)} & C_{(\nu)|(\alpha)}, D_{(\lambda)| (m)\rho}

\nonumber

\end{array} \end{eqnarray}
}

In the case where the fields satisfy trace constraints, we have simply
to require the $C$ and $D$ fields and all their derivatives in the
above table to be traceless fields, while the condition
$\varphi^\prime=2D$ 
is implemented by removing the components $\varphi_{11}$ and all their
derivatives from the $y_A$. The Cauchy order is unchanged.  

\section{Killing tensors in flat space}

Consider real functions $\Lambda(a^{\dag\mu},x^\mu)$, where the
dependence of $\Lambda$ on the variables $a^{\dag\mu}$ is assumed to
be polynomial, while the dependence on $x^\mu$ is assumed to be
smooth. Let us introduce the operators
\begin{eqnarray}
S^\dag = a^{\dag\mu}\frac{\d}{\d x^\mu} \ ,\ 
N_a=a^{\dag\mu}\frac{\d}{\d a^{\dag\mu}}\ , \ 
\Box=\eta^{\mu\nu} \frac{\d}{\d x^{\mu}}\frac{\d}{\d x^{\nu}}\ ,\  
S= \eta^{\mu\nu} \frac{\d}{\d a^{\dag\mu}}\frac{\d}{\d x^\nu}.
\end{eqnarray}
The gauge parameters for the models of level $s-1$ can then be
identified with functions $\Lambda_{s-1}$ satisfying $N_a 
\Lambda_{s-1}=(s-1)\Lambda_{s-1}$ and equations (\ref{16}) are
equivalent to 
\begin{equation}
S^\dag \Lambda_{s-1} = 0 \ , \
S \Lambda_{s-1}=0\ , \ \Box \Lambda_{s-1}=0\ .\label{eq fond 1}
\end{equation}

Introducing the additional operators
\begin{eqnarray}
\bar S^{\dag}= x^\mu\frac{\d}{\d a^{\dag\mu}} \ , N_x =
x^\mu\frac{\d}{\d x^\mu} \ , \ H = N_a - N_x \ ,
\end{eqnarray}
the subset $S^\dag,\bar S^\dag,H$ form an \emph{sl(2)} algebra, 
\begin{eqnarray}
[S^\dag,\bar S^\dag]=H\ ,\ 
[h,S^\dag] = 2S^\dag \ , \ 
[H,\bar S^\dag] = - 2\bar S^\dag \ .
\end{eqnarray}
represented on polynomials in $a^\dagger$ with coefficients in smooth
functions of $x$ with $S^\dag$ and $\bar S^\dag$ acting as creation
and destruction operators respectively. The results now follow from
standard arguments used to describe $sl(2)$ representations which we
spell out explicitly in order to be self-contained.

Because they involve exactly $s-1$ variables $a^{\dag\mu}$, the
parameters $\Lambda_{s-1}$ also satisfy the condition
\begin{equation}
(\bar S^\dag)^k \Lambda_{s-1} = 0 \ , \ \forall k\geq s\\ . \label{con
ini}
\end{equation}
The solutions $\Lambda_{s-1}$ to $S^\dag\Lambda_{s-1}=0$ can then be
classified according to the lowest integer $0\leq m \leq s-1$ such that 
$(\bar S^\dag)^{s-m} \Lambda_{s-1} = 0$. Applying $S^\dag$ to the
latter equation, one can easily show using the commutation relations
that 
\begin{equation}
\Big[(s-m)(s-m-1)(\bar S^\dag)^{s-m-1} +
(s-m)H (\bar S^\dag)^{s-m-1}\Big]\Lambda_{s-1} =
0 \ . \label{rel}
\end{equation}
Using the definition of $H$, this relation gives 
\begin{eqnarray}
  \label{eq:15}
  N_x\Big((\bar S^\dag)^{s-m-1}\Lambda_{s-1}\Big)=(s-1)(\bar
  S^\dag)^{s-m-1}\Lambda_{s-1}\,,
\end{eqnarray}
and thus $N_x \Lambda_{s-1}=m\Lambda_{s-1}$. The gauge
parameters of the different classes of solutions characterized by the
integer $0\leq m \leq s-1$ are thus of homogeneity $m$ in
$x^\mu$. Injecting $\Lambda_{s-1}(x,a^\dag)=a^{\dag\mu_1}\dots a^{\dag
  \mu_{s-1}} A_{\mu_1\dots\mu_{s-1}|\nu_1\dots \nu_m}x^{\nu_1}\dots
x^{\nu_m}$ in the equation 
$S^\dag\Lambda_{s-1}=0$ then implies that the constants
$A_{\mu_1\dots\mu_{s-1}|\nu_1\dots \nu_m}\Lambda_{s-1}$ have the
symmetries of the Young tableau
\begin{equation} \label{eq:YT22}
  \begin{picture}(100,25)
    \put(0,-10){
      \put(-12,23){\tiny $\mu$}
      \put(-12,13){\tiny $\nu$}
      \put(0,30){\line(1,0){90}}
      \put(0,20){\line(1,0){90}}
      \put(90,20){\line(0,1){10}}
      \put(80,20){\line(0,1){10}}
      \put(63,21.5){$\cdots$}
      \put(60,20){\line(0,1){10}}
      \put(50,10){\line(0,1){20}}
      \put(0,10){\line(1,0){50}}
      \put(40,10){\line(0,1){20}}
      \put(16.5,21.5){$\cdots$}
      \put(16.5,11.5){$\cdots$}
      \put(10,10){\line(0,1){20}}
      \put(0,10){\line(0,1){20}}
    }
  \end{picture}
\end{equation}
where the first row is of length $s-1$ and the second row is of length
$0\leq m \leq s-1$. 

That $S \Lambda_{s-1}=0$ requires the tableaux to be traceless in the
rows is now obvious, while the commutation relation $[S,S^\dag]=-\Box$
implies that $\Box \Lambda_{s-1}=0$ is a consistency condition for
$S^\dag \Lambda_{s-1}=0 =S\Lambda_{s-1}=0$.  .




\providecommand{\href}[2]{#2}\begingroup\raggedright\endgroup

\end{document}